\documentclass[conference]{IEEEtran}
\IEEEoverridecommandlockouts
\usepackage{verbatim}
\usepackage{longtable}
\usepackage{adjustbox}
\usepackage{amsmath}
\usepackage{amsfonts}
\usepackage{amssymb}
\usepackage{multirow}
\usepackage{amsmath,amssymb,amsfonts}
\usepackage{fancyhdr}
\usepackage{graphicx}
\usepackage{textcomp}
\usepackage{xcolor}
\usepackage{enumitem}
\usepackage{float}
\usepackage{graphicx}
\usepackage[normalem]{ulem}
\usepackage{url}
\useunder{\uline}{\ul}{}
\usepackage{cite}
\usepackage{amsmath,amssymb,amsfonts}
\usepackage{graphicx}
\usepackage{textcomp}
\usepackage{comment}
\usepackage{subcaption}
\usepackage{caption}
\usepackage{xcolor}
\def\BibTeX{{\rm B\kern-.05em{\sc i\kern-.025em b}\kern-.08em
    T\kern-.1667em\lower.7ex\hbox{E}\kern-.125emX}}
\usepackage{algorithm}
\usepackage[noend]{algpseudocode}

\begin{document}

\title{Collaborative Proof-of-Work: A Secure Dynamic Approach to Fair and Efficient Blockchain Mining}

\author{\IEEEauthorblockN{Rizwanul Haque$^{1}$, SM Tareq Aziz$^{2}$, Tahrim Hossain$^{3}$, Faisal Haque Bappy$^{4}$,\\ Muhammad Nur Yanhaona$^{5}$, and Tariqul Islam$^{6}$}
\IEEEauthorblockA{
$^{1, 2}$ University of Dhaka,
$^{3, 4, 6}$ Syracuse University, and
$ ^{5}$ BRAC University\\
Email: mdrizwanul-2018925355@cs.du.ac.bd, smtareq-2018825356@cs.du.ac.bd, \\ mhossa22@syr.edu, fbappy@syr.edu, nur.yanhaona@bracu.ac.bd, and mtislam@syr.edu} 
}

\maketitle

\thispagestyle{fancy}
\lhead{\small{This work has been accepted at the 2025 IEEE 15th Annual Computing and Communication Workshop and Conference (CCWC 2025)}}
\cfoot{}

\begin{abstract}

Proof-of-Work (PoW) systems face critical challenges, including excessive energy consumption and the centralization of mining power among entities with expensive hardware. Static mining pools exacerbate these issues by reducing competition and undermining the decentralized nature of blockchain networks, leading to economic inequality and inefficiencies in resource allocation. Their reliance on centralized pool managers further introduces vulnerabilities by creating a system that fails to ensure secure and fair reward distribution. This paper introduces a novel Collaborative Proof-of-Work (CPoW) mining approach designed to enhance efficiency and fairness in the Ethereum network. We propose a dynamic mining pool formation protocol that enables miners to collaborate based on their computational capabilities, ensuring fair and secure reward distribution by incorporating mechanisms to accurately verify and allocate rewards. By addressing the centralization and energy inefficiencies of traditional mining, this research contributes to a more sustainable blockchain ecosystem.
\end{abstract}


\section{Introduction}
Blockchain technology, has transformed industries by enabling decentralized and secure platforms. Central to blockchain’s functionality are consensus algorithms, which allow independent nodes within the network to achieve agreement on transaction validity, block inclusion, and order without the need for a central authority. By reaching consensus, these algorithms ensure data integrity, resilience, and accuracy across the blockchain, safeguarding the network from errors or malicious attacks \cite{bashir2019blockchain}. 

Proof of Work (PoW) is one of the earliest and most widely used consensus algorithms in blockchain renowned for its robust security \cite{10.1145/3579845}. It secures networks like Bitcoin \cite{nakamoto2009bitcoin} by requiring miners to solve cryptographic puzzles to validate transactions and add new blocks, ensuring that tampering or fraudulent activity is prohibitively costly. Miners compete to find a specific nonce that, when combined with the current block’s header, generates a hash value below a set difficulty target. This process aligns miners’ interests with the network’s integrity through block rewards and transaction fees, incentivizing honest participation. Despite its strengths, PoW also presents significant challenges. Its reward structure disproportionately favors strong miners with greater computational resources, consolidating mining power and undermining decentralization \cite{huang2021rich}. This centralization enables exploitative behaviors \cite{10.1145/2976749.2978341}. Additionally, weak miners often waste computational power as they are less likely to win rewards, leading to inefficiencies and substantial energy waste.

Concentrated mining power in PoW systems increases vulnerabilities, such as selfish mining and block withholding attacks, where dominant miners can exploit their position to disrupt operations and gain unfair rewards \cite{feng2019selfish, cheng2021evolutionary}. Additionally, the substantial energy demands required to secure PoW networks raise concerns about scalability and environmental impact, emphasizing the importance of developing more energy-efficient approaches \cite{sedlmeir2020energy, li2019energy}.
To address challenges in efficiency, decentralization, security and fairness recent work explores innovative solutions such as partial contribution rewards \cite{szalachowski2019strongchain}, low-difficulty transactions \cite{10.1145/3087801.3087809}, energy repurposing for productive tasks \cite{li2020}, and decentralized mining pools \cite{luu2017smartpool}. While these approaches offer promising improvements, they come with shortcomings, including misaligned incentives that weaken security, heightened vulnerability to DoS attacks from increased network overhead, centralization risks due to reliance on task managers, and scalability constraints arising from high gas fees and operational costs in blockchain networks. These limitations highlight the need for innovative frameworks that balance efficiency, fairness, and decentralization. To overcome these significant issues, in this work, we focused on the following three research questions.

\textit{\textbf{RQ1:} How can we optimize computational resources through dynamic mining pool formation in Collaborative Proof-of-Work (CPoW) systems while ensuring decentralized collaboration among miners?} 

\textit{\textbf{RQ2:} What strategies can ensure equitable reward distribution in CPoW models for miners with lower capabilities, without compromising incentives for stronger miners?} 

\textit{\textbf{RQ3:} How can we enhance energy efficiency in mining operations within the CPoW framework, and what trade-offs exist between energy consumption and performance?}

To address these questions, this work introduces a novel Collaborative Proof-of-Work framework, which proposes a dynamic mining pool architecture. Unlike static pools, CPoW enables miners to collaborate flexibly, sharing both computational power and rewards equitably inside the network, thereby promoting decentralization. The main contributions of this work are:

\begin{itemize}
    \item \textbf{Dynamic Mining Pool Formation.} We proposed a flexible protocol that allows miners to form or leave pools dynamically based on their computational capabilities, ensuring adaptability and resource optimization.
    
    \item \textbf{Fair And Secure Reward Distribution.} We designed a shared reward system that ensures equitable profit distribution among miners, while incorporating mechanisms to verify contributions and facilitating the secure transfer of rewards.

    \item \textbf{Theoretical Foundation and Security Analysis.} We establish a rigorous theoretical framework for the Collaborative Proof-of-Work (CPoW) model, integrating key concepts, assumptions, and mechanics with a comprehensive security evaluation.
    
    \item \textbf{Energy Efficiency.} Our approach can reduce the overall energy consumption of mining by distributing computational workloads more effectively through collaboration.
\end{itemize}

In the remainder of this paper, we review related work in Section II, focusing on existing collaborative mining approaches and their limitations. In Section III, we present the formal foundation, where we define the key concepts and models for our framework. Section IV explores the system workflow of our Collaborative Proof of Work (CPoW) model. In Section V, we conduct a security analysis, followed by a performance evaluation in Section VI. Finally, in Section VII, we conclude with insights on the framework’s implications and outline future directions.

\section{Related Work}
Proof of Work (PoW) consensus mechanisms, widely used in blockchain networks, face critical issues regarding the fairness of block reward distribution, security against exploitative mining behaviors, and the centralization of mining power. 

Hunag et al. \cite{huang2021rich} investigate the disproportionate acquisition of block mining rewards by large mining entities in PoW systems, resulting in a "rich get richer" dynamic. The authors explain that miners with greater computational power have a higher probability of successfully mining blocks, allowing them to capture more rewards consistently. This cycle concentrates rewards among large miners, undermining decentralization as smaller miners are discouraged from participating, further centralizing power.

Studies by Feng et al\cite{feng2019selfish} and Cheng et al. \cite{cheng2021evolutionary} illustrate that concentrated mining power heightens vulnerabilities in PoW systems, increasing susceptibility to selfish mining and block withholding attacks. When a small number of miners control a significant share of the network’s mining power, they can strategically withhold blocks to disrupt network operations and enhance their own rewards at the expense of smaller miners.

The substantial energy demands of PoW systems, which secure networks through intensive computation, are highlighted by Sedlmeir et al. \cite{sedlmeir2020energy} and further discussed by Li et al. \cite{li2019energy}. These demands raise scalability and environmental concerns, emphasizing the need for more energy-efficient approaches.

StrongChain introduces "weak solutions" to reward partial contributions in PoW, reducing energy waste and leveling the playing field for smaller miners \cite{szalachowski2019strongchain}. However, this approach departs from PoW’s emphasis on full solutions, potentially weakening its security guarantees and encouraging miners to prioritize partial rewards, leading to unintended incentives.

The FruitChains protocol, proposed by Pass and Shi, addresses issues like mining variance and selfish mining by introducing “fruits”, transactions with lower mining difficulty that can be appended to blocks mined at higher difficulty \cite{10.1145/3087801.3087809}. This approach aims to reduce reliance on mining pools by offering more frequent rewards. However, FruitChains introduces new challenges, as low difficulty fruits can lead to high transmission overhead and may even open the network to denial-of-service attacks. Additionally, duplicate fruits are discarded, leading to inefficiencies and increased reward variance. These limitations highlight the complexities involved in modifying traditional Proof of Work (PoW) protocols.

Li et al. \cite{li2020} introduce the Proof of Neural Architecture Search (PoNAS) consensus, which improves mining efficiency by redirecting computational power from repetitive hash calculations, traditionally used in Proof of Work (PoW), to neural architecture search (NAS) tasks. Instead of solely expending energy on cryptographic puzzles, PoNAS allocates this computational effort toward finding optimal neural network architectures. This shift allows mining energy to contribute both to blockchain security and to advancements in deep learning, addressing PoW’s high energy consumption by giving it a dual purpose. However, centralization risks are introduced by relying on a mining pool manager to allocate tasks and manage rewards, creating a single point of control and compromising the blockchain’s decentralized nature.

SmartPool is a decentralized mining pool implemented on Ethereum that eliminates the need for a central operator by using smart contracts to manage shares and distribute rewards \cite{luu2017smartpool}. Miners submit shares, which are verified probabilistically to reduce gas costs, and rewards are distributed based on these verified shares. While SmartPool enhances decentralization, it faces limitations from Ethereum gas fees and potential scalability issues.

Collectively, these studies highlight significant advancements but reveal a critical gap: existing solutions often prioritize specific aspects such as fairness, energy efficiency, or decentralization, without achieving a unified framework that balances these goals while maintaining robust security and scalability.

\section{Formal Foundation}
Proof of Work (PoW) relies on cryptographic hash functions to validate blocks in a blockchain. For a block to be accepted as valid, its hash must be lower than a specific threshold set by the network's difficulty level, meaning it must begin with a certain number of leading zeros. The key to achieving this lies in finding a suitable nonce, an arbitrary number that, when combined with the block header and hashed, produces a hash meeting these criteria. In PoW systems, miners must search through a large range of possible nonces, \( N_{\text{total}} \), repeatedly generating and testing hashes until they find one that meets the requirement. The time \( T_{\text{solo}} \) required by a single miner to exhaust the nonce range \( N_{\text{total}} \) depends on their computational power \( C_{p} \). Since \( C_{p}^{\text{weak}} \ll C_{p}^{\text{strong}} \), this implies \( T_{\text{solo}}^{\text{weak}} \gg T_{\text{solo}}^{\text{strong}} \). In other words, miners with lower computational power require significantly more time to complete the search than those with higher computational power. 

Given that \( C_{p}^{\text{weak}} \ll C_{p}^{\text{strong}} \), the probability \( P_{\text{success}}^{\text{weak}} \) of a weak miner successfully solving the PoW puzzle remains extremely low compared to \( P_{\text{success}}^{\text{strong}} \). As a result, weak miners will earn rewards \( R_{\text{weak}} \approx 0 \) over time, making mining economically unsustainable for them. This causes mining power to become concentrated among strong miners, centralizing control within a few powerful entities. This situation contradicts blockchain's core principle of decentralization, which aims to distribute control and rewards across a wide range of participants. In a decentralized network, all miners should have a fair chance at earning rewards. Concentration of rewards among a few powerful miners undermines this openness and inclusivity.

This would lead weaker miners to collaborate and form groups in an effort to compete. Multiple weak miners can form a collaborative group \( G_{\text{collab}} \). In this setup, miners within \( G_{\text{collab}} \) divide the nonce search range \( N_{\text{total}} \) into smaller sub-ranges.

Let, \( C_{p}^{m_i} \) denote the computational power of miner \( m_i \). The time \( T_{\text{solo}}^{m_i} \) required for a single weak miner \( m_i \) to exhaust \( N_{\text{total}} \) without collaboration is given by:
\[
T_{\text{solo}}^{m_i} = \frac{N_{\text{total}}}{C_{p}^{m_i}}
\]
In a collaborative mining setup, each miner \( m_i \in G_{\text{collab}} \) is assigned a portion of the range \( N_i = \frac{N_{\text{total}}}{|G_{\text{collab}}|} \). The time \( T_{\text{collab}} \) required for the group \( G_{\text{collab}} \) to collectively exhaust \( N_{\text{total}} \) becomes:
\[
T_{\text{collab}} = \frac{N_i}{C_{p}^{m_i}} = \frac{N_{\text{total}}}{|G_{\text{collab}}| \cdot C_{p}^{m_i}}
\]
For a group of \( n \) weak miners, the time savings can be represented as

\[
T_{\text{collab}} \approx \frac{T_{\text{solo}}^{m_i}}{n}
\]

In a static collaborative mining setup, groups are fixed, with set participants and roles, limiting flexibility. By contrast, dynamic collaborative mining allows miners to join or leave \( G_{\text{collab}} \) as needed, providing greater flexibility. We develop a framework that supports this dynamic collaborative mining, enabling miners to cooperate efficiently and on-demand. This approach increases the probability of successfully mining a block, maximizes rewards, and minimizes wasted computational effort. However, dynamic collaborative mining presents several key challenges, including establishing trust among participants by verifying individual contributions and ensuring fair and secure reward distribution within \( G_{\text{collab}} \). These challenges are discussed in detail below.

\subsection{Trust Dynamics in Contribution Verification}
In a collaborative mining group \( G_{\text{collab}} \), each miner \( m_i \) is expected to contribute a portion of computational power \( C_{p}^{m_i} \) toward the total computational power \( C_{p}^{G_{\text{collab}}} = \sum_{i=1}^{|G_{\text{collab}}|} C_{p}^{m_i} \). This combined power determines the group’s likelihood of successfully mining a block, where the probability of success is \( P_{\text{success}}^{G_{\text{collab}}} \propto C_{p}^{G_{\text{collab}}} \). Each miner should ideally receive a reward \( R_{m_i} \propto \frac{C_{p}^{m_i}}{C_{p}^{G_{\text{collab}}}} \). In a dynamically formed collaborative mining group \( G_{\text{collab}} \), each miner faces uncertainty regarding the actual computational contributions of others. Without a trusted authority or prior verification, freeriders \( f_j \in G_{\text{collab}} \) could potentially claim rewards without contributing substantial \( C_{p}^{f_j} \). This situation creates a scenario where
\[
C_{p}^{f_j} \approx 0 \quad \text{but} \quad R_{f_j} > 0
\]
This affects the entire group’s effective computational power \( C_{p}^{G_{\text{collab}}} \). The presence of freeriders results in
\[
C_{p}^{G_{\text{collab}}} = \sum_{i=1}^{|G_{\text{collab}}|} C_{p}^{m_i} - \sum_{j=1}^{|F|} C_{p}^{f_j}
\]
where \( F \subset G_{\text{collab}} \) denotes the subset of freeriders. Consequently, the group's mining success probability \( P_{\text{success}}^{G_{\text{collab}}} \) is reduced.
\subsection{Challenges in Reward Distribution}

Ensuring fair and secure reward distribution remains a critical challenge as collaborative mining evolves from static to dynamic frameworks. While static collaborative mining setups rely on centralized management, dynamic systems must address similar issues but in a decentralized manner. In a static setup, when a block is successfully mined by the group, rewards are typically sent to a single, central etherbase address \( E \). This etherbase address \( E \) is controlled by the pool operator, who holds the private key \( k_E \) necessary to access and distribute the rewards. However, the pool operator controlling \( k_E \) introduces key issues:
\begin{itemize}
    \item \textbf{Reliance on Operator's Integrity:} Participants must trust the operator to distribute \( R_{m_i} \) fairly. Without oversight, the operator could reduce rewards such that \( R_{m_i}' < R_{m_i} \), giving miners less than they are due.
    \item \textbf{Single Point of Failure:} If \( k_E \) is compromised, all rewards at \( E \) are at risk. Any loss or misuse of \( k_E \) could prevent miners from receiving \( R_{m_i} \), effectively reducing rewards to zero: \( R_{m_i} \approx 0 \quad \text{if } k_E \text{ is lost or misused.} \)
\end{itemize}
In the context of dynamic collaborative mining, the challenge is to address these issues without relying on a central operator to manage \( k_E \). Instead, a decentralized approach is required to ensure fair reward distribution and safeguard rewards, allowing miners to collaborate effectively.

\section{System Workflow}

\begin{figure*} [t]
    \centering
    \includegraphics[width=0.77\textwidth]
{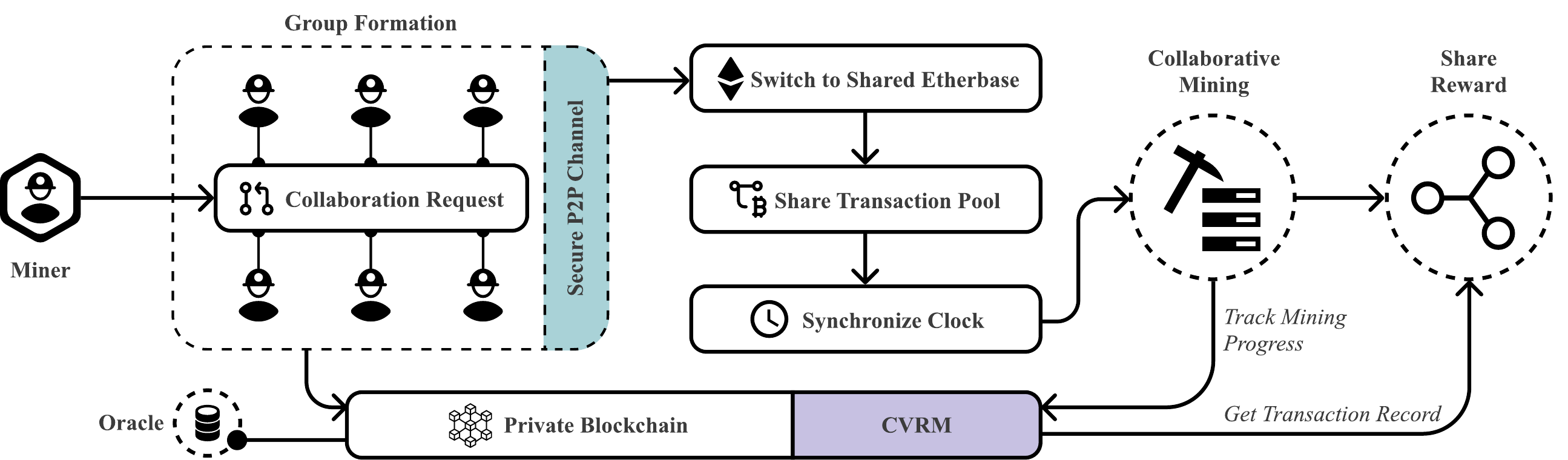}
    \caption{Collaborative Mining process in CPoW} 
    \label{fig:workflow} 
\end{figure*}
The Collaborative Proof of Work (CPoW) framework enables secure, fair, and efficient mining within a decentralized, collaborative environment by addressing key challenges in flexibility, fairness and trust. In CPoW, three core processes,  dynamic group formation, contribution management, and decentralized reward distribution work together to overcome the unique obstacles of dynamic collaborative mining.

Dynamic group formation solves the problem of flexibility by allowing miners to form well-balanced collaborative groups based on computational power. Contribution management ensures fairness by accurately tracking each miner's work, verifying their computational effort to ensure that only those who contribute benefit from the group's efforts, thereby preventing any undue advantage for freeriders. Finally, decentralized reward distribution removes the need for a central authority to manage and distribute rewards by enabling fair, direct, and secure allocation to each participant, eliminating the risk of withheld rewards and single point of failure.

This section presents a comprehensive overview of how these processes function together in the CPoW framework to create a dynamic collaborative mining environment as depicted in Figure \ref{fig:workflow}.

  
  
  
  
  
  


\subsection{Group Formation and P2P Setup}
The collaborative mining process in CPoW begins with group formation as shown in Fig.\ref{fig:workflow}. Initially, each node calculates its local hashrate using a dedicated module (Alg.\ref{alg:group_formation} lines 1-2). The hashrate represents each node’s computational power, specifically its ability to solve cryptographic puzzles by finding a suitable nonce a number that, when combined with the block header and hashed, produces a result below the network's difficulty target.  A node seeking collaboration broadcasts a request message to nearby nodes (Alg.\ref{alg:group_formation} lines 3-4). Upon receiving this request, neighboring nodes evaluate the sender’s hashrate and respond with an acceptance message if they find the hashrate compatible (Alg.\ref{alg:group_formation} lines 5-8). Nodes prioritize accepting the first collaboration request with a similar hashrate, ensuring a balanced and efficient mining group.

Once a group is formed, members establish peer-to-peer (P2P) connections and add the collaborating peer list to their trusted peer set, enhancing group connectivity (Alg.\ref{alg:group_formation} lines 9-10). To facilitate cooperation, group members switch their local etherbase, the account designated to receive mining rewards to a shared etherbase (Alg.\ref{alg:group_formation} line 11) as illustrated in Fig. \ref{fig:workflow}. The etherbase is the address where mining payouts are sent. The switch to a shared etherbase ensures rewards can be pooled and fairly distributed among all collaborators. 

As depicted in Fig.\ref{fig:workflow} a shared transaction pool is established for collaborative mining, while each node’s local pool remains active to handle new transactions independently (Alg.\ref{alg:group_formation} line 12). Transactions from the local pool are ordered within the shared pool, which uses the Ricart-Agrawala mutual exclusion algorithm \cite{ricart1981optimal} to manage the insertion of transactions in a coordinated manner. This mechanism prevents conflicts by ensuring mutual exclusion during transaction handling. To address potential clock synchronization issues, collaborators share their timestamps prior to mining, as outlined in Fig. \ref{fig:workflow}, and employ Berkeley's algorithm to compute a synchronized timestamp that is used in the block being mined (Alg.\ref{alg:group_formation} line 13).

The group then proceeds to divide the nonce range between the collaborators. The total range is divided equally among collaborators, with a slight overlap between assigned ranges for contribution verification purposes. Each miner searches within their designated range to find the desired hash using the Dagger Hashimoto \cite{dagger} algorithm (Alg.\ref{alg:group_formation} lines 14 - 15).
\begin{algorithm}
  \caption{Group Formation and Collaborative Mining}\label{alg:group_formation}

  \begin{algorithmic}[1] 
    \For{each $node \in net$}
        \State $node.\texttt{calcHash}()$
        \If{$node.\texttt{reqCollab}()$}
            \State $node.\texttt{sendReq}(neighbors)$
            \For{each $nbr \in neighbors$}
                \If{$nbr.\texttt{matchHash}(node)$}
                    \State $nbr.\texttt{acceptReq}()$
                    \State $group.\texttt{add}(nbr)$
                \EndIf
            \EndFor
        \EndIf
        \If{$group.\texttt{ready}()$}
            \State $peers \gets group.\texttt{connectPeers}()$
            \State $group.\texttt{setSharedEthBase}()$
            \State $group.\texttt{sharePool}(peers)$
            \State $group.\texttt{syncClocks}(peers)$
            \State $range \gets group.\texttt{splitRange}(peers)$
            \State $group.\texttt{startMining}(range)$
    so    \EndIf
    \EndFor
  \end{algorithmic}
\end{algorithm}
\subsection{Contribution Management with CVRM}
\begin{algorithm}
  \caption{Contribution Management with CVRM}\label{alg:contribution_management}
  \begin{algorithmic}[1] 
    
    \State $group.\texttt{startDKG}()$
    
    \For{each $m \in group$}
        \State $share \gets m.\texttt{generateShare}()$
        \State $\texttt{CVRM.storeShare}(m, share)$
    \EndFor
    
    \State $group.\texttt{pubKey} \gets \texttt{DKG.getPublicKey}()$
    \State $group.\texttt{etherbase} \gets \texttt{DKG.getEtherbase}()$
    
  \end{algorithmic}
\end{algorithm}

To securely manage mining activities, we employ a \textbf{private blockchain} system featuring the Contribution Verifier and Reward Manager (CVRM) as shown in Fig. \ref{fig:workflow}.During group formation, collaborators initiate a Distributed Key Generation (DKG) protocol \cite{gennaro2003secure} to create individual private key shares. Each member registers their private key share with the CVRM (Alg.\ref{alg:contribution_management} lines 1-4). Threshold cryptographic algorithms \cite{stathakopoulou2017threshold} enable secure group operations that require only a subset of key shares. These key shares collectively generate a shared public key (Alg.\ref{alg:contribution_management} line 5). From this shared public key, a shared etherbase address is derived, serving as the collective account for mining rewards (Alg.\ref{alg:contribution_management} line 6). This approach allows secure and decentralized management of mining rewards, ensuring that control is shared among collaborators. It enhances trust, transparency, and reliability, allowing rewards to be distributed fairly without relying on any single member.
\subsection{Reward Tracking and Distribution}

We utilize an oracle to monitor etherbase addresses continuously, collecting data on balances and activities (Alg.\ref{alg:reward_tracking} line 1) as illustrated in Fig. \ref{fig:workflow}. When it is time to distribute rewards, each collaborator is required to submit proof of their contribution to the CVRM (Alg.\ref{alg:reward_tracking} lines 2-3). This proof consists of a mapping of nonce to hash values they calculated during mining. This mapping allows the CVRM to verify each collaborator’s work by cross-checking the overlapping portions of nonce ranges between collaborators. If the hashes from the overlapping sections match between collaborators, the proof is accepted, and the collaborators are allowed to withdraw their rewards. However, if discrepancies are detected in the submitted proofs, the CVRM recalculates the hashes for the overlapping nonces to identify inconsistencies. This recalculation helps the CVRM determine the honest collaborators by verifying whose proof aligns with the expected hash outputs (Alg.\ref{alg:reward_tracking} line 4). 

\begin{algorithm}
  \caption{Reward Tracking and Distribution}\label{alg:reward_tracking}
  \begin{algorithmic}[1] 
    
    \State $\textit{oracle}.\texttt{monitorEtherbase}(group.\texttt{etherbase})$
    
    \For{each $c \in group$}
        \State $proof \gets c.\texttt{generateProof}()$
        \If{$\texttt{CVRM.verify}(proof)$}
            \State $rewardAmount \gets \texttt{CVRM.reward}(c)$
            \State $\texttt{CVRM.withdraw}(c, rewardAmount)$
        \EndIf
    \EndFor
    
  \end{algorithmic}
\end{algorithm}

Once the proof of contribution has been verified, each collaborator submits a transaction specifying the address to which they wish to transfer their rewards. The CVRM then reviews the transaction details, signs the transaction with its authority key, and returns the signed transaction to the collaborator (Alg.\ref{alg:reward_tracking} lines 5-6). This signed transaction allows the collaborator to securely transfer their reward to the specified address, ensuring that the distribution process is both authenticated and controlled by the CVRM, maintaining the integrity and fairness of the reward allocation.

\begin{figure*}[htbp]
  \centering
  \captionsetup{justification=centering}
  \begin{subfigure}[b]{0.24\textwidth}
    \centering
    \captionsetup{justification=centering}
    \includegraphics[width=\textwidth]{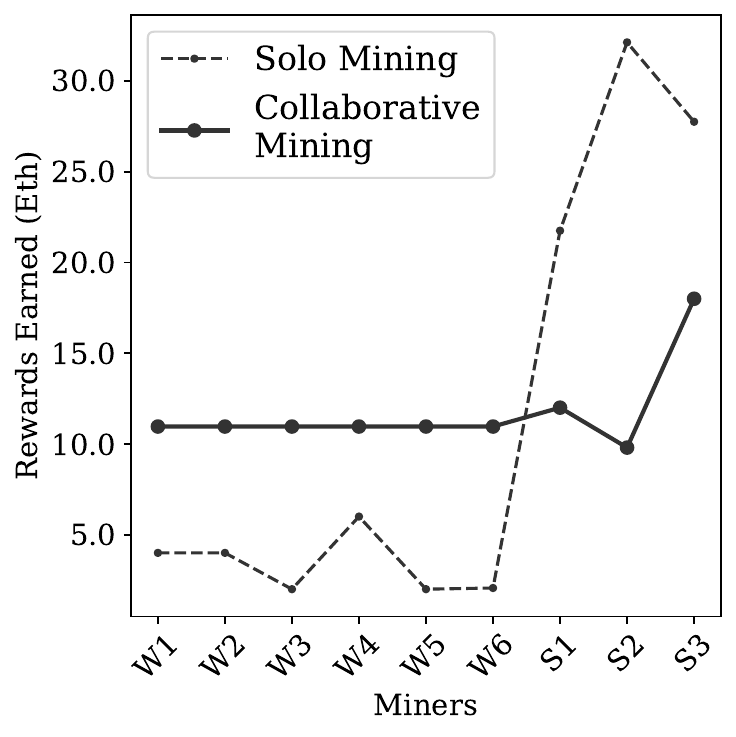}
    \caption{Reward Distribution\\(mining 50 blocks)} 
    \label{fig:reward}
  \end{subfigure}
  \hfill
  \begin{subfigure}[b]{0.24\textwidth}
    \centering
    \captionsetup{justification=centering}
    \includegraphics[width=\textwidth]{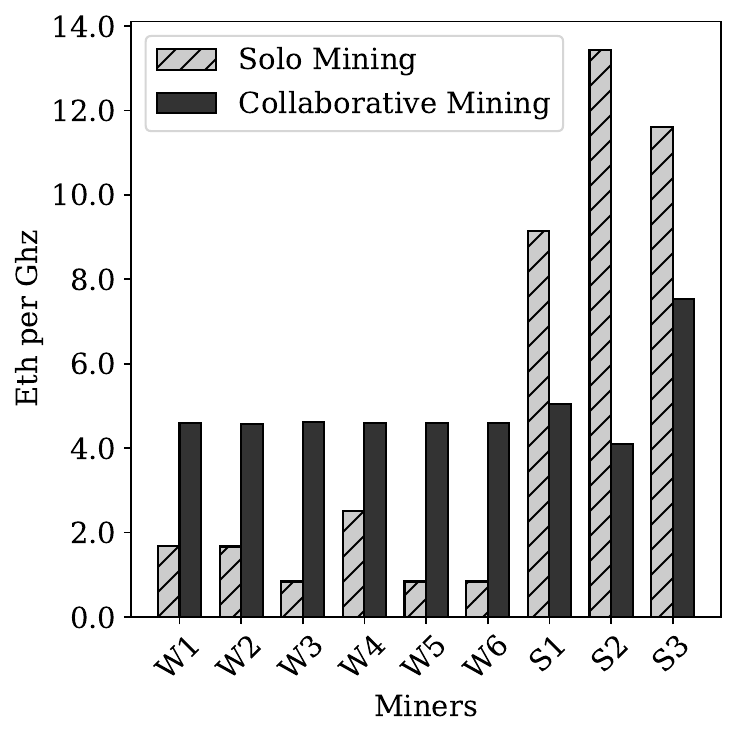}
   \caption{Mining Efficiency Comparison (Eth/GHz)}    \label{fig:eth-ghz}
  \end{subfigure}
  \hfill
  \begin{subfigure}[b]{0.24\textwidth}
    \centering
    \captionsetup{justification=centering}
    \includegraphics[width=\textwidth]{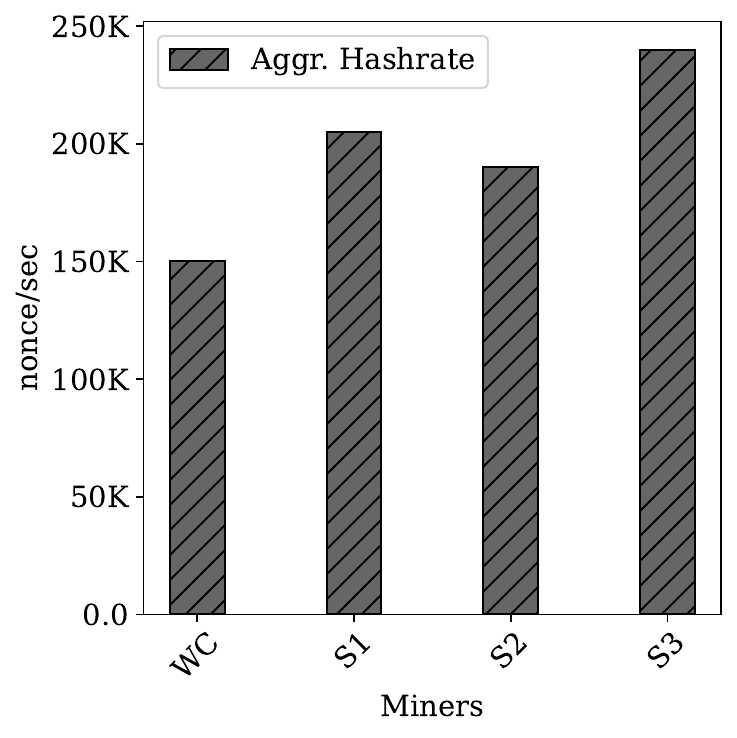}
    \caption{Aggregated Hashrate of Weak Miners} 
    \label{fig:agg}
  \end{subfigure}

  \caption{Performance Comparison between Solo and Collaborative Mining}
  \end{figure*}
  
\section{Security Analysis}

In the collaborative mining framework of CPoW, secure and fair reward distribution is critical. The Contribution Verifier and Reward Manager (CVRM) plays a central role by verifying contributions and ensuring that only honest collaborators receive rewards in a secure manner. This section demonstrates the framework’s resilience against dishonest behavior and its effectiveness in preserving reward integrity.
\subsection{Analysis of Contribution Verification Process}
The verification process uses nonce-to-hash mappings and overlapping nonce ranges between collaborators to maintain reward integrity and promote honesty. By cross-checking these overlaps, the Contribution Verifier and Reward Manager (CVRM) can efficiently verify contributions without recalculating costly hash functions. Since collaborators don’t know with whom they share overlaps, nor the specific overlapping nonce range collusion is virtually impossible, ensuring secure and fair reward distribution.

The effectiveness of this verification mechanism is formalized in the following statement, which demonstrates how the contribution verification mechanism helps preserve both reward integrity and collaborative honesty. An analysis follows to confirm these properties.

\textit{\textbf{Claim:} Given a set of collaborators \( C = \{ c_1, c_2, \dots, c_n \} \), if each collaborator \( c_i \) submits a proof of contribution \( p_i \) verified by the CVRM, then only honest participants \( C_{\text{honest}} = \{ c_i \in C \mid p_i \text{ is valid} \} \) are allowed to access rewards.}

\textbf{Analysis:} Each collaborator \( c_i \) has a nonce range \( N_i \), where \( N_i = \{ n_{i,1}, n_{i,2}, \dots n_{i,n} \} \). The proof \( p_i \) submitted by \( c_i \) includes a mapping of each \( n_{i,j} \in N_i \) to its corresponding hash \( H_{i,j} = f(n_{i,j}) \), where \( f \) is a cryptographic hash function. This hash function is applied to each nonce in an attempt to solve the cryptographic puzzle essential for mining a successful block.

To verify the accuracy of each \( p_i \), the CVRM checks overlapping nonce ranges \( N_{i,j} = N_i \cap N_j \) between two collaborators \( c_i \) and \( c_j \).

For each pair of collaborators \( c_i \) and \( c_j \) with an overlapping nonce range \( N_{i,j} = N_i \cap N_j \), the CVRM verifies that

\[
H_i(x) = H_j(x) \quad \forall x \in N_{i,j}
\]

Here, \( H_i(x) = f(x) \) and \( H_j(x) = f(x) \) are the hash values submitted by \( c_i \) and \( c_j \) for each \( x \in N_{i,j} \). If this condition holds for all \( x \in N_{i,j} \), the CVRM validates the contributions of both collaborators.

If a collaborator \( c_i \) submits a falsified proof \( p_i \), discrepancies in the hashes for overlapping nonces \( N_{i,j} \) will reveal inconsistencies. In cases of discrepancy, the CVRM recalculates the expected hash values for \( N_{i,j} \) to determine the correct hashes \( H_{\text{true}} \), identifying the dishonest collaborator. Therefore, any collaborator with an invalid proof \( p_i \) is excluded from \( C_{\text{honest}} \) and is denied access to rewards.
\subsection{Analysis of Secure Reward Distribution Framework}
Building on the contribution verification process, secure reward distribution is crucial to maintaining the integrity of the Collaborative Proof of Work (CPoW) framework.
In the collaborative mining framework, secure reward distribution is achieved through controlled access to the private key \( K \). The CVRM ensures that \( K \) is only reconstructed for verified participants, preventing unauthorized access. This process is formalized as follows:

\textit{\textbf{Claim:} Let \( C_{\text{eligible}} = \{ c_i \in C \mid p_i \text{ is valid} \} \) denote verified collaborators. The CVRM reconstructs \( K \) only when \( \forall c_i \in C_{\text{eligible}}, \; p_i \text{ is valid} \). No collaborator or CVRM component holds \( K \) independently.}

\textbf{Analysis:} \( K \) exists in distributed shares \( K_1, K_2, \dots, K_m \) across the collaborators \(c_i\). \( K \) is formulated as
\[
K = \text{Combine} (K_1, K_2, \dots, K_m)
\]
by the CVRM only for signing after verification. If \( C_{\text{eligible}} \neq \emptyset \), then \( K \) is temporarily reconstructed and used to sign reward transactions \( T_i \) for each \( c_i \in C_{\text{eligible}} \). Unauthorized reward access is prevented since \( K \) is inaccessible outside verified signing, ensuring rewards \( R_i \) are distributed only to \( C_{\text{eligible}} \). Together, these mechanisms (verification and controlled key access) uphold the integrity and fairness of the reward distribution process.

\section{Performance Evaluation}
To assess the performance of our proposed collaborative mining process, we conducted a series of experiments in a controlled environment, implementing our approach on the Ethereum (Geth version 1.11.6 \cite{GEthereum}) network. The configuration included Ubuntu 22.04 (LTS) on a 64-bit architecture with an AMD Ryzen 5 3550H CPU, with a base speed of 2.10 GHz, 4 cores, and 8 threads. A total of nine nodes were set up, with six designated as weak miners and three as strong miners. To simulate the weak miners, we introduced a 5 ms delay before verifying each nonce value.


\subsection{Reward Distribution Analysis:} In solo mining, weak miners earned an average of 3.34 ETH, with considerable variability (standard deviation of 1.62), indicating inconsistent earnings (Figure \ref{fig:reward}). However, in collaborative mining, their average reward increased to 10.96 ETH, with much lower variability (standard deviation of 0.01). This shift suggests that collaborative mining offers a more equitable distribution of rewards. 

Strong miners also showed a change, earning an average of 11.40 ETH in solo mode but only 5.04 ETH in collaborative mining. This indicates that while strong miners earned less collaboratively, the overall fairness improved for all miners.

\subsection{Efficiency Comparison:} We also calculated Ethers per GHz to assess mining efficiency (Figure \ref{fig:eth-ghz}). Weak miners achieved an average of 1.4 Ethers per GHz in solo mining, while in collaborative mining, their efficiency rose significantly to 4.6 Ethers per GHz, reflecting consistent performance among participants. Strong miners’ efficiency decreased from 13.94 Ethers per GHz in solo mining to 6.0 Ethers per GHz in the collaborative setting, highlighting a shift in how resources were allocated. 

\subsection{Hashrate Analysis:} In our configuration, strong miners have about eight times the hashrate of weak miners. Strong miners achieve around 200,000 nonces per second, while weak miners have less than 30,000 nonces per second. Despite only three strong miners, they contribute 81.3\% of the total computation power, while weak miners contribute just 19.7\%. This centralization limits the chances for weak miners to successfully mine a block, presenting a challenge to their participation. 

Combining the hashrates of weak miners brings it closer to that of an individual strong miner (Figure \ref{fig:agg}). This significantly enhances the chances of weak miners successfully mining a block. Collaboration of weak miners proves to be an effective strategy, enhancing their overall contribution to the mining process.





\begin{table*}[]
\caption{Feature Comparison With Existing Approaches}
\label{tab:comp}
\resizebox{\textwidth}{!}{%
\begin{tabular}{cllllllll}
\hline
\multicolumn{1}{l}{\textbf{}} &
  \textbf{\begin{tabular}[c]{@{}l@{}}Dependency on\\ Pool Manager\end{tabular}} &
  \textbf{\begin{tabular}[c]{@{}l@{}}Dependency on\\ Strong Miners\end{tabular}} &
  \textbf{\begin{tabular}[c]{@{}l@{}}Reward \\ Security\end{tabular}} &
  \textbf{\begin{tabular}[c]{@{}l@{}}Resource\\ Requirements\end{tabular}} &
  \textbf{Scalability} &
  \textbf{\begin{tabular}[c]{@{}l@{}}Subspace \\ Overlapping\end{tabular}} &
  \textbf{\begin{tabular}[c]{@{}l@{}}Prob. of \\ Wasted Effort\end{tabular}} &
  \textbf{Cost} \\ \hline
\textbf{\cite{mihaljevic2022approach}} &
  Yes &
  No &
  \begin{tabular}[c]{@{}l@{}}No (Distributed by \\ Pool Miners)\end{tabular} &
  High &
  Low &
  No &
  No &
  High \\ \hline
\textbf{\cite{LI2022100089}} &
  Yes &
  Yes &
  \begin{tabular}[c]{@{}l@{}}No (Distributed by \\ Pool Miners)\end{tabular} &
  No &
  Low &
  No &
  No &
  No Extra Cost \\ \hline
\textbf{SmartPool \cite{luu2017smartpool}} &
  No &
  No &
  Yes &
  \begin{tabular}[c]{@{}l@{}}Heavy dependency\\ on smart contracts\end{tabular} &
  Scalable &
  Yes &
  Yes &
  High \\ \hline
\textbf{P2Pool \cite{p2pool}} &
  No &
  Yes &
  \begin{tabular}[c]{@{}l@{}}No (Pool’s Reward \\ address set manually)\end{tabular} &
  ShareChain &
  Low &
  Yes &
  Yes &
  No Extra Cost \\ \hline
\textbf{\begin{tabular}[c]{@{}c@{}}Our\\ Approach\end{tabular}} &
  No &
  No &
  Yes &
  No &
  Scalable &
  No &
  No &
  No Extra Cost \\ \hline
\end{tabular}%
}
\end{table*}\

\subsection{Feature Comparison with Existing Approaches}
Table \ref{tab:comp} presents a comprehensive comparison of our approach against existing solutions, evaluating eight critical features that impact the overall effectiveness and practicality of mining pool implementations. Our approach demonstrates several significant advantages over existing solutions. Unlike traditional approaches \cite{mihaljevic2022approach} and \cite{LI2022100089}, our system eliminates dependency on pool managers, reducing centralization risks. 

In terms of security, our approach matches SmartPool \cite{luu2017smartpool} in providing robust reward security, while avoiding the vulnerabilities of solutions \cite{mihaljevic2022approach}, \cite{LI2022100089}, and P2Pool, which rely on potentially risky distribution mechanisms through pool miners or manual reward address configurations. Unlike SmartPool’s reliance on smart contracts and P2Pool’s ShareChain requirements, our design minimizes resource consumption without sacrificing security. It matches SmartPool’s scalability while overcoming the limitations and subspace overlapping issues that affect \cite{mihaljevic2022approach}, \cite{LI2022100089}, and P2Pool, ensuring more efficient use of computational resources. Cost-wise, it maintains the "No Extra Cost" advantage of \cite{LI2022100089} and P2Pool while avoiding the high operational expenses of \cite{mihaljevic2022approach} and SmartPool.

\section{Conclusion}
In this paper, we introduced a dynamic collaborative mining approach to enhance fairness and efficiency within the Ethereum network. Our collaborative PoW allows for dynamic mining pool formation and optimizes energy consumption by sharing workloads, addressing the environmental concerns of traditional PoW systems. This framework promotes greater decentralization and inclusivity by enabling miners with lower computational power to participate effectively, reducing the dominance of high-powered entities in blockchain mining. Its fair reward distribution ensures equitable compensation for weaker miners, fostering economic fairness and wider participation in the blockchain ecosystem. Overall, this research has broad implications for improving energy efficiency, equity, and inclusivity in decentralized systems.
\bibliographystyle{IEEEtran}
\bibliography{IEEEabrv,references}
\end{document}